\documentclass[lettersize,journal]{IEEEtran}
\usepackage{authblk}
\usepackage{amsmath,amsfonts}
\usepackage{algorithmic}
\usepackage{array}
\usepackage{cuted}
\usepackage{amsmath}
\usepackage{soul}
\usepackage{hyperref}

\usepackage[usenames,dvipsnames]{xcolor}

\usepackage{textcomp}
\usepackage{stfloats}
\usepackage[longend,ruled,linesnumbered]{algorithm2e}
\usepackage[font=footnotesize,labelfont=bf]{caption}
\usepackage{subcaption}
\usepackage{mathtools}

\usepackage{stfloats}
\usepackage{url}
\usepackage{verbatim}
\usepackage{graphicx}
\def\BibTeX{{\rm B\kern-.05em{\sc i\kern-.025em b}\kern-.08em
    T\kern-.1667em\lower.7ex\hbox{E}\kern-.125emX}}
\usepackage{balance}
\usepackage{array,multirow,graphicx}
\newcolumntype{L}{>{\centering\arraybackslash}m{1.8cm}}

\title{Multi-Agent Reinforcement Learning with Action Masking for UAV-enabled Mobile Communications}
\author{Danish Rizvi}
\author{David Boyle}
{\vspace{-2mm}}
\affil{Imperial College London, UK}

{\vspace{-5mm}}

\begin{document}
\maketitle
\RestyleAlgo{ruled}

\begin{abstract}
Unmanned Aerial Vehicles (UAVs) are increasingly used as aerial base stations to provide \textit{ad hoc} communications infrastructure. Building upon prior research efforts which consider either static nodes, 2D trajectories or single UAV systems, this paper focuses on the use of multiple UAVs for providing wireless communication to mobile users in the absence of terrestrial communications infrastructure. In particular, we jointly optimize UAV 3D trajectory and NOMA power allocation to maximize system throughput. Firstly, a weighted K-means-based clustering algorithm establishes UAV-user associations at regular intervals. The efficacy of training a novel Shared Deep Q-Network (SDQN) with action masking is then explored. Unlike training each UAV separately using DQN, the SDQN reduces training time by using the experiences of multiple UAVs instead of a single agent. We also show that SDQN can be used to train a multi-agent system with differing action spaces. Simulation results confirm that: 1) training a shared DQN outperforms a conventional DQN in terms of maximum system throughput (+20\%) and training time (-10\%); 2) it can converge for agents with different action spaces, yielding a 9\% increase in throughput compared to mutual learning algorithms; and 3) combining NOMA with an SDQN architecture enables the network to achieve a better sum rate compared with existing baseline schemes.

\end{abstract}

\begin{IEEEkeywords}
 Disaster management, trajectory optimization, non-orthogonal multiple access, unmanned aerial vehicle, reinforcement learning
\end{IEEEkeywords}
\section{Introduction}
\IEEEPARstart{A}{major} step in post-disaster recovery is the deployment of rescue teams at appropriate locations. Moreover, effective disaster management response entails uninhibited communication of rescue teams with the Disaster Control Center (DCC). In scenarios with compromised terrestrial communication infrastructure, use of UAVs as aerial base stations (ABS) is being widely advocated in the scientific community owing to advantages such as swift deployment, high mobility and better probability of providing line-of-sight (LoS) communications~\cite{7470933}. UAVs can thus be used to aid, or temporarily replace, terrestrial networks for providing a critical communication channel between DCC and rescue teams \cite{8641424}.

Despite increased popularity for disaster scenarios, the use of UAVs for communication poses serious challenges including optimal 3D trajectory and resource allocation design. One of the major hurdles for practically deploying UAVs as ABS in disaster scenarios is to satisfy constraints such as limited energy and number of connected users~\cite{8315206}. UAV communications networks are typically time-bound due to the limited available power onboard. Therefore, an efficient transmission allocation mechanism is required to maintain a robust and long-lasting communication network. Moreover, an effective 3D trajectory design scheme is also essential to enhance user coverage, increase data rate and minimize interference from nearby UAVs.
In a bid to tackle the problem of inhibited communication during a disaster scenario, techniques like the positioning of permanent infrastructures such as micro base stations and 5G hotspots have been proposed~\cite{6736747}. However, compared to such schemes, UAV networks can provide enhanced flexibility in the sense that the size and distribution of a UAV fleet can vary adaptively based on the density and distribution of users. Similarly, from an economic perspective, benefits reaped from UAV-based networks outweigh those of micro base stations insofar as the latter, while acting as permanent access points, are difficult to dismantle and reposition when the services are no longer required or required elsewhere. On the contrary, the redeployment of UAVs may be far more efficient, which aligns nicely with the concept of green communications.
{\vspace{-2mm}}
\subsection{State-of-the-art}

Acknowledgment of the use of UAVs as a potential solution for providing emergency communications has paved way for numerous research efforts. A summary of most relevant research works discussed in this section is provided in Table \ref{literature_review}. Topics of interest subsumed in this research fall under the umbrella of the following UAV network aspects:

\subsubsection{Network Topology and UAV Trajectory Design}

Numerous recently proposed solutions suffer with regard to practicality and comprehensiveness considering the manifold variables involved in an actual disaster scenario. In~\cite{9416239} the authors focus on secrecy provisioning in UAV-ground communication while considering both uplink and downlink communications. However, an LoS-based channel model is assumed instead of a more realistic probabilistic LoS model. Similarly, researcher in \cite{8663615} also consider pure LoS wireless channels. This issue has been arrested in \cite{9034493}, which invokes the probabilistic LoS channel model for UAV-aided data collection in scenarios with a dominant shadowing effect (e.g., low flying UAVs). Research in \cite{9762672} focuses on maximizing average data rates from ground nodes by optimizing trajectory of multiple UAVs but also uses a TDMA scheme. In \cite{7510870}, the authors aim to minimize total transmit power by optimizing UAV hovering location but employ FDMA. However, TDMA and FDMA (part of Orthogonal Multiple Access) schemes both suffer from a reduced spectral efficiency as compared to NOMA scheme \cite{8269066}.

Design of UAV trajectory, especially for a multi-UAV system, is a challenging task. Authors in \cite{8913466} and \cite{9151993} endeavor to optimize 2D trajectory of a single UAV by minimizing the total mission time and the average age of information, respectively. Solutions that aim to optimize 2D trajectory often neglect that the area of coverage and service quality can be improved by varying UAVs’ altitudes. In \cite{8663350}, the authors propose a convex optimization method to determine best hover position and power allocation policy for a single UAV. Solutions aimed at single UAVs inherently fail to address inter-UAV interference when exposed to a multi-UAV environment and serve to motivate our work. In \cite{9529207}, the authors consider multiple static UAVs for servicing non-orthogonal multiple access (NOMA) users and optimizing their 3D deployment. The authors in \cite{8254236,8269066} employ a NOMA scheme to provide enhanced connectivity to users. However, both of these works employ a circular trajectory for the UAV fleet where users can access one UAV after another and is not optimal for cases where most of the users being served are sparsely located or stationary. Nevertheless, circular trajectory serves as one useful benchmark for our work.

\subsubsection{Non-orthogonal Multiple Access (NOMA) in UAV networks}
The NOMA technique has allowed networks to gain better spectral efficiency and increased system throughput \cite{8114722}. Despite an increasing number of publications, research on NOMA maintains room for advancement. In \cite{8648498}, the authors compare OMA and NOMA techniques for multiple UAV scenarios and maximize throughput by joint optimization of coverage radii and power allocation for static ground users only. In~\cite{tabassum2016non}, the authors investigate the increase of achievable spectral efficiency, but cluster all users into a single NOMA group, which considerably increases the complexity and decreases the reliability of the successive interference cancellation (SIC)~\cite{8114722}. Other works assume to limit the maximum number of NOMA users per channel as N = 2, e.g.,~\cite{8632657,9177252}, to meet QoS requirements, reduce co-channel interference, and simplify SIC at receivers. Such solutions, in fact, achieve problem simplification at the cost of decreased accuracy of the solution \cite{9903854}. NOMA clustering techniques proposed in \cite{8788625} cater to larger cluster sizes but use heuristic approaches that lead to higher computational costs and scalability issues. To reduce clustering complexity, the authors in \cite{9520121} invoke the K-means clustering algorithm for UAV-user association and then apply NOMA for cellular offloading in a congested downlink network setup. However, in a bid to train a single Deep Q Learning (DQN) for all UAVs, the authors fix the number of users that can be connected to all UAV clusters, which again is a simplification of the problem at the cost of accuracy. A key limitation of such an architecture is the inability to cope with users either joining or leaving the clusters during service, as UAVs are not trained to serve varying cluster sizes. This limitation motivates our work to allow UAVs to service clusters with dynamically varying sizes.

\subsubsection{Joint Optimization and Reinforcement Learning (RL) in UAV-aided networks}
Many research efforts point out the utility of RL in optimizing the numerous aspects of UAV-aided networks such as 2D and 3D trajectory design \cite{10016705}, NOMA power allocation \cite{9860079}, clustering for UAV-user association \cite{8269066} and interference suppression \cite{hassan2022interference}. However, many recent efforts either optimise trajectory of a single UAV, e.g.,~\cite{9517120,10039197}, or deployment of multiple static UAVs, \cite{9164901,9817159,9310353,9817159}, to serve ground nodes. Motivated by this, our solution focuses on optimum resource allocation for mobile ground users through a mobile UAV fleet.

Research efforts in \cite{9817159,9310353} use separate DQNs for all agents to learn independently in a multi-UAV environment. This is usually slow since DQNs have only their own experience to learn from and the performances across different agents may thus vary substantially. The authors in \cite{9817159} propose an ensemble RL framework consisting of multiple DQNs. The DQNs run in parallel in a bid to maximize sum rate in a dynamic environment and varying QOS requirements for ground nodes at the cost of increased computation. To cater this, one of the benchmarks used in this research \cite{9520121}, proposes a mutual RL approach in which all UAVs share their experience to train a single DQN and find the best UAV trajectory and power allocation scheme. The authors demonstrate that experience sharing can can result in better convergence time than separately trained DQNs and motivates our research.
\begin{table*}
\captionsetup{width=.85\linewidth}
\caption{Summary of the most relevant recent research highlighting major limitations in three key areas: \textbf{i)} Topology and Trajectory (implementations suited for scenarios with single mobile user, multiple static users, UAV 2D trajectory design, single UAV 3D trajectory design or use of OMA), \textbf{ii)} NOMA (fixed cluster size, computationally intensive user clustering) and \textbf{iii)} Reinforcement learning (Seperate DQNs for agents, fixed No. of users).} 
\setlength{\tabcolsep}{0.2em}
    \centering
    \begin{tabular}{ |>{\centering}m{1.6cm} | >{\centering}m{3cm} | >{\centering}m{2cm}| >{\centering}m{1.3cm}|  >{\centering}m{1.3cm}| >{\centering}m{1.3cm}|  >{\centering}m{1.3cm}| L|} 
    
        \hline
        \textbf{Ref} & \textbf{Objective} &\textbf{Realistic Channel Model} & \textbf{3D UAV Trajectory} &  \textbf{Multiple UAVs} & \textbf{Nodes Mobility} & \textbf{Shared Learning} & \textbf{Dynamic cluster sizes} \\
        \hline
        \multicolumn{8}{|c|}{\textbf{Network Topology and UAV Trajectory}} \\
        \hline    
        
        Y. Zeng et al.~\cite{8663615}& Minimize total transmit and flight power  & \textbf{x}& \textbf{x} & \textbf{x} & \textbf{\checkmark} & \textbf{x} & \textbf{x}\\
        \hline
        C. You et al.~\cite{9034493}& Maximize minimum mean data rate  & \textbf{\checkmark} & \textbf{\checkmark}& \textbf{x} & \textbf{x} & \textbf{x} & \textbf{x} \\
        \hline
        M. Mozaffari et al.~\cite{7510870}& Minimize total transmit power  & \textbf{x}& \textbf{x} & \textbf{\checkmark} & \textbf{x} & \textbf{x} & \textbf{x} \\
        \hline
        P. Sharma et al.~\cite{8269066} & Minimize outage and maximize data rate  & \textbf{\checkmark} & \textbf{x}& \textbf{x} & \textbf{x} & \textbf{x} & \textbf{x}\\
        \hline
        W. Shi et al.~\cite{9529207}& Minimize transmit power and maintain QoS  & \textbf{\checkmark} & \textbf{x}& \textbf{\checkmark} & \textbf{x} & \textbf{x} & \textbf{x} \\
        \hline
        J. Lyu et al.~\cite{8254236} & Minimize outage and maximize data rate  & \textbf{x}& \textbf{\checkmark} & \textbf{\checkmark} & \textbf{\checkmark} & \textbf{x} & \textbf{\checkmark} \\
        \hline
       
        \multicolumn{8}{|c|}{\textbf{Non-Orthogonal Multiple Access}} \\
        \hline
        Y. Sun et al.~\cite{8648498}& Maximize throughput  & \textbf{\checkmark}& \textbf{\checkmark} & \textbf{\checkmark}  & \textbf{x} & \textbf{x} & \textbf{x} \\
        \hline
        S. Khairy et al.~\cite{9177252} & Maximize system capacity  & \textbf{\checkmark}& \textbf{x} & \textbf{\checkmark} & \textbf{x} & \textbf{x} & \textbf{x} \\
        \hline
        H. Nguyen et al.~\cite{8788625} & Max spectral efficiency and user fairness  & \textbf{\checkmark} & \textbf{x}& \textbf{x} & \textbf{x} & \textbf{x} & \textbf{\checkmark} \\
        \hline
        \multicolumn{8}{|c|}{\textbf{Reinforcement Learning}} \\
        \hline
        S. Zhou et al.~\cite{9762672} & Max average transmission rate  & \textbf{x} & \textbf{\checkmark}& \textbf{\checkmark} & \textbf{\checkmark} & \textbf{x} & \textbf{x} \\
        \hline
         R. Zhong et al.~\cite{9520121}&Sum rate maximization&  \textbf{\checkmark}& \textbf{\checkmark} & \textbf{\checkmark} & \textbf{\checkmark} & \textbf{\checkmark} & \textbf{x} \\
         \hline
        J. Wang et al.~\cite{9164901}& Minimize sum power with user QoS  & \textbf{\checkmark} & \textbf{x}&\textbf{\checkmark}& \textbf{x} & \textbf{x} & \textbf{x} \\
        \hline
         S. Mahmud et al.~\cite{9817159}& Sum rate maximization & \textbf{\checkmark} &\textbf{x}& \textbf{\checkmark} & \textbf{x} & \textbf{x} & \textbf{x} \\
        \hline
          \multicolumn{8}{|c|}{\textbf{This Work}}  \\
        \hline
    
        - & Sum rate maximization & \textbf{\checkmark} & \textbf{\checkmark} & \textbf{\checkmark} & \textbf{\checkmark} & \textbf{\checkmark} & \textbf{\checkmark} \\
        
        \hline
    
    \end{tabular}

\label{literature_review}
\vspace{-4mm}
\end{table*}
\vspace{-2mm}

\subsection{Motivation}
The existing literature provides the foundations for solving various challenges in UAV-enabled networks, in particular contributing towards 2D and 3D deployment and trajectory design of UAVs for optimal communications. However, a large proportion of literature focuses on solutions with static ground nodes~\cite{9310353}. In such cases, UAVs act as a temporary base station, as the optimal deployment position reduces to a fixed 3D co-ordinate. Design of 3D trajectory and NOMA power allocation schemes for dual-ended (UAVs and users) mobility gets increasingly difficult with use of conventional approaches (matching theory, convex optimization, and game theory). Moreover, this mobility renders dynamic user clustering and design of 3D trajectory an NP-hard problem. Therefore, when tested against such challenges, existing algorithms designed for resource management and UAV deployment are exposed to issues such as high complexity and static assumptions, highlighted in \cite{8743390} and \cite{9348645}. Furthermore, employment of NOMA and 3D trajectory design to maximize UAV network parameters in disaster-based mobility models has not been addressed adequately \cite{8121998}.

For a highly complex case such as deployment of multi-UAV system in highly dynamic environments, machine learning approaches are able to use past experience to offer a near-optimal solution. The main reason for employing RL is that it can orchestrate a promising solution by exploring and interacting with dynamic environments, obtaining rewards for multiple solutions, and utilizing high-reward experiences~\cite{7967963}. However, most solutions based on RL propose independent model training for all agents, which are unable to benefit from other agents’ experiences.

This research also aims to answer questions related to shared model learning between all UAV agents in which agents share their experiences to catalyze the training process. With more experiences to learn from, this approach reduces the network convergence time. As discussed above, the authors in \cite{9520121} gather and exploit the experiences of multiple UAVs to train a single DQN. However, in order to share the same model between agents, the author suggests an equal distribution of users amongst UAVs. We demonstrate that such forced allocation of an equal number of users to all UAVs results in sub-optimal sum rates. Moreover, we propose an algorithm that allows the system to make clusters spontaneously and improve system throughput.\vspace{-3mm}

\subsection{Contributions}
In this work, we make the following specific contributions:
\begin{itemize}
\item We propose a novel NOMA-based framework for 3D deployment of a multi-UAV system to serve downlink communication to terrestrial vehicular users. Based on the proposed system model, we jointly optimize the trajectory of a UAV fleet and user power allocation.
Keeping in view user mobility, we show that dynamic clustering improves system sum data rate (27\%) and how user speed affects the system throughput.
\item A two-step solution is proposed to the problem. Firstly, to cater to UAV location during clustering, we periodically invoke the weighted K-means clustering algorithm. Once users are associated with UAVs, we propose the using a novel SDQN algorithm to optimize throughput by regularly assessing an optimal 3D trajectory and power allocation policy. To the best of our knowledge, this work is the first to allow shared learning amongst UAVs with non-overlapping action spaces. Our solution also allows UAVs to cater to dynamically changing cluster sizes during service by exploiting shared experiences, thereby allowing users to leave and join the network anytime. We show that the proposed SDQN algorithm outperforms the leading conventional DQN algorithm~\cite{9817159} in terms of convergence time by 10\%. 
\item Simulation results\footnote{Supplementary communication efficiency analyses concerning UAV flight energy are also made available online \url{https://github.com/smrizvi1/UAV_energy_analysis_SDQN}. Moreover, all code and research artefacts will be made openly available upon acceptance of the manuscript for publication.} presented in this paper demonstrate that: 1) the proposed NOMA-based framework offers a higher achievable sum-rate as compared to OMA (FDMA); 2) the proposed SDQN, through experience sharing, has a shorter convergence time compared to independent learning through separate DQNs; 3) Optimal 3D trajectory design through the proposed framework outperforms circular deployment and 2D trajectory; 4) Timely re-clustering is a mandatory part of the proposed framework and helps to optimize UAV-user association when ground users diverge geographically; and 5) Invoking the proposed SDQN architecture with invalid action masking (Section \ref{secIV}) yields 9\% higher throughput than comparable mutual learning schemes, e.g.,~\cite{9520121}. 
\end{itemize}

\begin{table*}
\caption{Notations}
    \centering
    \begin{tabular}{|c| c| c| c|} 
        \hline
        Noun & Notation & Noun & Notation \\
        \hline
        UAV fleet size & $U$ & Number of users & $K$ \\  
        \hline
        Service duration & $T$ & Re-clustering period & $T_r$\\
        \hline
        Maximum user speed & $V_{max}$ & Probability of LoS / NLos & $P_{LoS}/P_{NLoS}$\\
        \hline
        LoS/ NLoS Path loss & $L_{LoS}/ L_{NLoS}$ & UAV $u$ to user $k$ path loss & $l^u_k$\\
        \hline
        UAV $u$ to user $k$ 3D distance & $d_k^u$ & UAV $u$ to user $k$ channel gain & $g_k^u$\\
        \hline
        UAV $u$ to user $k$ fading co-efficient & $H_k^u$ & Power allocated by UAV $u$ to user $k$ & $P_k^u$\\
        \hline
        Equivalent channel gain & $G_k^u$ & Serving indicator & $v_u^k$\\
        \hline
        Carrier frequency& $f_c$ & AWGN& $\sigma$\\
        \hline
        Decoding order & $\pi$ & Bandwidth & $B$\\
          \hline
        SINR & $\gamma$ & Acheivable data rate & $\mathcal {R}$\\
        \hline
        user cluster& $C$ & Cluster vector mean& $\mu$\\
          \hline
        location of user $k$ & $l_k$ & maximum allowable users with UAV & $\upsilon$\\
      \hline
        Current state& $S$ & Action & $A$\\
          \hline
        Next state& $S'$ & Reward& $R$\\
          \hline
        Function for Q-value & $Q()$ & Discount Factor& $\beta$\\
          \hline
        Neural network parameters & $w$ &Greedy co-efficient& $\epsilon$ \\
          \hline
        Learning Rate& $\alpha$ & Target value for DQN & $y$\\
          \hline
        Loss function& $J()$ & Memory replay size& $e$\\
          \hline
    \end{tabular}

\vspace{-2mm}
\end{table*}

\section{SYSTEM MODEL}
\subsection{System Description}
\begin{figure}[!htb]

     \centering
     \includegraphics[width=3.5in]{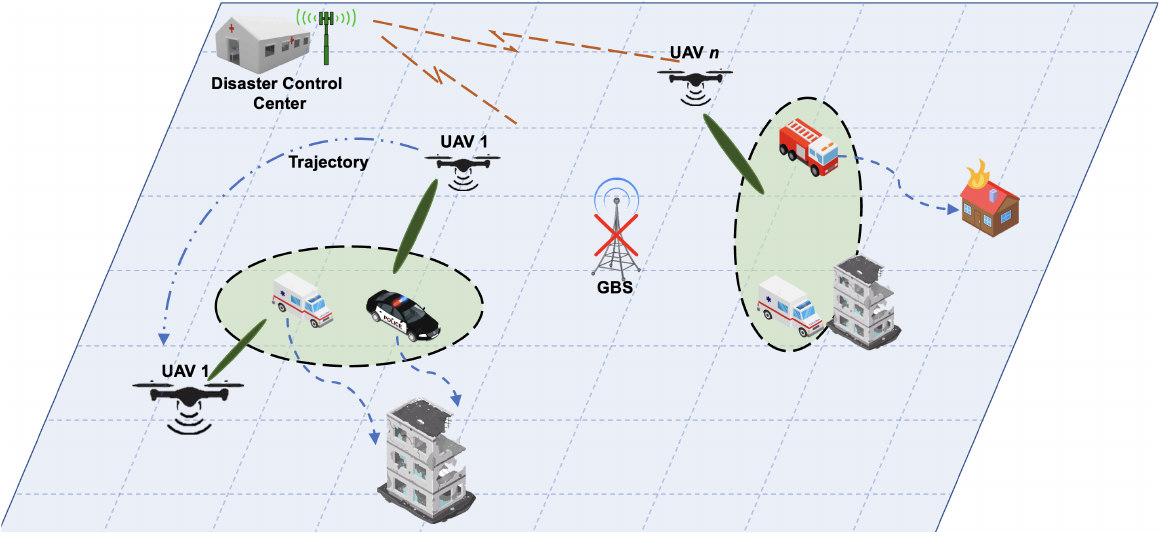}
     \caption{NOMA-assisted UAV network framework for Disaster recovery}\label{framework_uav_network}

   \vspace{-4mm}
\end{figure}

As shown in Fig.~\ref{framework_uav_network}, we consider an outdoor downlink communication setup in a disaster scenario with no central Ground Base Station (GBS) and numerous mobile and static rescue teams. A multi-UAV-based communication framework is proposed to provide connectivity between DCC rescue teams. Each UAV functions as an ABS, is equipped with an isotropic antenna to radiate uniformly in all directions and employs the NOMA scheme. In contrast to users served using the OMA technique, NOMA users in a cluster communicate on a single frequency. This results in intra-cell interference, which is inhibited using Successive Interference Cancellation (SIC) - a technique fundamental to NOMA~\cite{8114722}. Furthermore, for this multi-cell network, it is assumed that all UAVs employ the same frequency to simplify handover. Using the same frequency, however, introduces inter-cell interference among users. The set of users served by a UAV is denoted as $U \in \mathbb {U}=\{1,2,3\ldots M\}$. All users are partitioned into \textit{K} cells, as per the user-UAV association, where $k \in \mathbb {K}=\{1,2,3\ldots N\}$. It should be noted that the number of cells always equals the number of deployed UAVs. Every user in cell \textit{k} can only be served by UAV \textit{k}. Multiple orthogonal resource blocks can be used by a single UAV to serve more than one cluster simultaneously. However, without loss of generality, this research assumes that each UAV serves only one cluster.

\subsection{Mobility Model}

As in the case of vehicles moving in a disaster-hit area, geographical models like the Manhattan grid~\cite{8170804} restrict mobile nodes from visiting the whole simulation area. On the other hand, combinations of spatial and temporal dependencies, such as obstacle placement and node priorities, in node movements add realism to the mobility model \cite{mahiddin2021review}. During simulations, it is imperative to select a mobility model which is designed to mimic the movements of rescue teams through the streets of a disaster-struck area. Therefore, we use a Manhattan grid model with spatial dependencies given that such a hybrid approach better fits our scenario and helps evaluate solution performance with realistic approximations.

The disaster area map, simulated as shown in Fig. \ref{UAV trajectory plan view}, uses a Manhattan map model along with Dijkstra's Algorithm to find a route between CCC (origin) and target destination. Vehicles can only choose specified routes to reach their respective destinations. Blocks of buildings are simulated and are not available for vehicles to pass through. The maximum speed of vehicles is $V_{\text {max}}$. Every coordinate box $(x,y)$ on the map has a speed cost, $C_{x,y}$, which functions in analogy to debris randomly dispersed on road in case of calamities such as tornadoes or earthquakes. As a result, the vehicle speed on any co-ordinate box is given by $V_u = V_{\text {max}} - C_{x,y}$ where $C_{x,y} \sim U(0,  7/10\cdot V_{\text {max}})$.
{\vspace{-2mm}}
\subsection{UAV-User Association}
The simulation is initialized with the partitioning of all users into \textit{u} clusters as per UAVs' and users' geographical locations. The clustering ensures that each user is associated and served by one UAV only. Additionally, user clustering as per spatial locations would help UAVs move closer to the cluster center in search of optimal channel gains, thereby reducing inter-cluster interference.

\textit{\noindent{Remark 1:} It is expected that mobile users, especially rescue teams originating from a disaster control center, will disperse to reach their respective destinations. In such a case, the initial clustering of users and their association with UAVs will most likely lose relevance as time passes. This motivates periodic re-clustering which is considered essential to form rational clusters and maintain optimal datarates.}

In view of the performance dictates of \textit{Remark1}, re-clustering of users is done periodically after time interval ${T_r}$. This divides the total serving time $\textit{T}$ in to ${T / T_r}$ clustering instances.

\subsection{Propagation Model}
The 3GPP specification release 15 provides the channel model between UAVs and the users \cite{3gpp20173gpp}. Line-of-sight (LoS) and non-line-of-sight (NLoS) link states govern the path loss ${L_{{\text {LoS/NLoS}}}}$, which is expressed as \eqref{1} provided on next page. Here ${h_{c}(t)}$  denotes the height of UAV \textit{u}, ${f_c}$ denotes the carrier frequency used for transmission, and ${d_k^u}$ represents the 3D distance between user ${k}$ and UAV ${u}$ at instance $t$ and can be expressed as

\begin{figure*}[t]
\setcounter{equation}{0}

\begin{equation}\begin{aligned}
{P_{{\text {LoS}}}(t)} = L_{\mathrm{LoS / NLoS}}(t)=&\begin{cases} 30.9 + \log_{10}d_{k}^{u}(t)(22.5 - 0.5\log_{10}h_{u}(t)) + 20\log_{10}f_{c}, &  \text{ for LoS link} \\ \max\{{L_{LoS}, 32.4 + 20\log_{10}f_{c} + \log_{10}d_{k}^{u}(t)(- 7.6\log_{10}h_{u}(t) + 43.2}) , &  \text{ for NLoS link} \end{cases}\end{aligned}\tag{1}\label{1}
\end{equation}
\vspace{-6mm}
\end{figure*}

\setcounter{equation}{1}
\begin{equation} {d^{u}_k(t) = \sqrt { [x_k^u(t) - x_u(t)]^2 +[y_k^u(t) - y_u(t)]^2+ h_u^2(t) },\label{2}}
\end{equation}

For propagation model, the LoS probability is denoted as ${P_{LoS}}$ and is given as \eqref{3} given at the bottom of the next page, where \linebreak ${d_0}= \text{max}[18, - 432.94+ 294.05 \cdot \log_{10}h_u(t)] $, whereas ${p_{1}}$ = $- 0.95 + 233.98\cdot {\log _{10}}{h_{u}}(t)$. Coherently, the probability of NLoS is ${P_{{\text {NLoS}}}}$ = 1 - ${P_{{\text {LoS}}}}$. Resultantly, the expected path loss, ${L_{k}^u(t)}$, between user ${k}$ and UAV ${u}$ for time instance ${t}$ is governed by

\begin{figure*}[b]
\setcounter{equation}{2}

\begin{equation}\begin{aligned}
\hline
 {P_{{\text {LoS}}}(t)} = \begin{cases} \displaystyle {1,}&{\text {if} ~ \sqrt {{{\left ({{d_{k}^{u}(t)} }\right)}^{2}} - {{\left ({{h_{u}(t)} }\right)}^{2}}} \leq {d_{0}},} \\ \displaystyle {\frac {d_{0}}{{\sqrt {{{\left ({{d_{k}^{u}(t)} }\right)}^{2}} - {{\left ({{h_{u}(t)} }\right)}^{2}}} }} + \exp \left \{{\frac {d_{0} -\sqrt {\left ({{d_{k}^{u}(t)} }\right)^{2}-\left ({{h_{u}(t)} }\right)^{2}}}{p_{1}} + }\right \},}&{\text {if} ~ \sqrt {{{\left ({{d_{k}^{u}(t)} }\right)}^{2}} - {{\left ({{h_{u}(t)} }\right)}^{2}}} > {d_{0}}} \end{cases}\label{3}\end{aligned}
\end{equation}

\setcounter{equation}{13}

\begin{equation}
    \begin{aligned}
{\gamma_{\pi(k)}^{u}(t) = \frac{{g_{\pi(k)}^{u}(t){v_{\pi(k),u}}(t)\sqrt{P^u_{\pi(k)}(t)}}}{{\sum \nolimits _{i=k+1}^{K^{u}}g_{\pi(i)}^{u}(t){v_{\pi(i),u}}(t)\sqrt{P^u_{\pi(i)}(t)}} + \sum \nolimits _{s = 1,s \ne u}^{U} {\sqrt {P^{s}(t)}g_{k}^{s}(t)+ {\sigma _{k}^{u}(t)} ^{2}}}}  \label{14} \!\!\\
    \end{aligned}
\end{equation}
\end{figure*}

\begin{equation*} 
L_{k}^{u}(t)=P_{\text {NLoS}} \cdot L_{\text {NLoS}} + P_{\text {LoS}} \cdot L_{\text {LoS}}.\tag{4}\label{4}
\end{equation*}

With consideration of small-scale fading, the channel gain between user $k$ and UAV ${u}$ at time instance ${t}$ is given by 

\begin{equation} 
{g_{k}^{u}(t) = \frac{{H_{k}^{u}}(t)} { {10^{{{  {0.1} \cdot {L^{u}_{k}}(t)} }}}}},\tag{5}\label{5}\end{equation}
where $H_k^u(t)$ is the fading coefficient, as described in \cite{8329013}, between UAV ${u}$ and user ${k}$.

\subsection{Signal Model}
\subsubsection{UAV transmitted / User received signals}We denote ${v_{u,k}}$ as the service indicator, with ${v_{u,k}}$ = 1 indicating that user ${k}$  is associated with, and being served, by UAV ${u}$. Otherwise, ${v_{u,k}}$ = 0. Resultantly, the superpositioned NOMA signal transmitted by UAV ${u}$ is given by  

\begin{equation*} 
{x^{u}(t) = \sum \limits _{k = 1}^{K} \sqrt {P_{k}^{u}(t)}\cdot x_{k}^{u}(t)\cdot{{v_{u,k}(t)}}  },\tag{6}\label{6}\end{equation*}

\noindent{}where ${x_{k}^{u}(t)}$ denotes the transmitting signal from UAV ${u}$ to user ${k}$ and ${P_{k}^{u}(t)}$ is the power allocated to user ${k}$. Combining equations \ref{5} and \ref{6} yields the received signal at user ${k}$ to be

\begin{equation*} 
{y_{k}^{u}(t) = g_{k}^{u}(t)x_{k}^{u}(t) + {I_{\text {inter}}}_{k}^{u}(t) + {I_{\text {intra}}}_{k}^{u}(t) + \sigma _{k}^{u}(t) },\tag{7}\label{7}\end{equation*}

\noindent{}where $\sigma _{k}^{u}(t)$ is the Additive White Gaussian Noise (AWGN),  ${I_{\text {inter}}}_{k}^{u}(t)$ represents the cumulative inter-cluster interference from all non-serving UAVs at user $k$, and $I_{\text {intra}_{k}}^{u}(t)$ represents interference within the cluster. 

\subsubsection{Intra and Inter-cell interference}

${I_{\text {inter}}}_{k}^{u}(t)$ can be described as

\begin{equation*} 
{{I_{\text {inter}}}_{k}^{u}(t) = \sum \limits _{s \ne u,s = 1}^{U} {x^{s}(t)\cdot g_{k}^{s}(t)\sqrt {P^{s}(t)}} },\tag{8}\label{8}\end{equation*}
\vspace{-4mm}

\noindent{}where ${g_{k}^{s}(t)}$ is the channel gain between user $k$ and UAV $s\neq u$, and $P^{s}(t)$ denotes the total transmission power of each of the non-serving UAVs $s\neq u$ and is given by  

\begin{equation*} 
{P^{s}(t) = \sum \limits _{k = 1}^{K} {{v_{u,k}}(t)P_{k}^{s}(t)} }.\tag{9}\label{9}\end{equation*}

Determining $I_{\text {intra}_{k}}^{u}(t)$ is a pre-requisite to finding the optimal decoding order and, subsequently, guarantee successful SIC, a technique which allows removal of intra-cluster interference and is executed at the receiver \cite{9043698}. For our case, channel gain of users and inter / intra-cluster interference changes at each instance due to mobility. Therefore, dynamic decoding order is used to extract signals at the user end. A supplementary term $G_{k}^{u}(t)$, which can also be regarded as equivalent channel gain, is introduced as a criterion to successfully determine the decoding order. It is expressed as 

\begin{equation*} {G_{k}^{u}(t) = \frac {{{v_{u,k}}(t)g_{k}^{u}(t)}}{{\sum \nolimits _{s = 1,s \ne u}^{T} {g_{k}^{s}(t)\sqrt {P^{s}(t)} + {{\sigma _{k}^{u}(t)} ^{2}}} }}}.\tag{10}\label{10}\end{equation*}

\noindent{}Since all users can change positions at all timeslots, dynamic decoding order is made effective by ascertaining it at each instance. Consider equivalent channel gains $G_{j}^{u}(t)$ and $G_{k}^{u}(t)$ for two users $j$ and $k$ respectively being served by UAV $u$. To remove signal of user $k$ at user $j$ using SIC, the equivalent channel gains must satisfy the following condition

\begin{equation*} G_{j}^{u}(t) \ge G_{k}^{u}(t),\tag{11}\label{11}\end{equation*}
derivable from \cite{8170332}. The inequality \eqref{11} implies that SIC has to be applied first at receivers with higher equivalent channel gain. 

Broadening the aforementioned scheme to a UAV $u$ serving $K^u$ users in a cluster at time slot $t$, the decoding order can be determined from comparing the equivalent channel gains of all users.  $G_{\pi (1) }^{u}(t) \le G_{\pi (2) }^{u}(t) \le \cdots \le G_{\pi (K^{u})}(t)$, where $\pi(k)$ represents the decoding order of all $k$. By this convention, $\pi(k)$ = 1 indicates that user 1 has the smallest equivalent channel gain and should be removed to get signal for user 2. Subsequently, SIC sequentially removes signals of all $(k-1)$ users to find the desired signal for user $k$. After SIC, the desired signal $S_{\pi (k)}$ and residual intra-cluster interference ${I_{\text {intra}}}_{\pi (k)}$ at user $k$ can be given as \eqref{12} and \eqref{13}, respectively.

\begin{equation}
 S_{\pi (k)}=x_{\pi (k)}^{u}(t)g_{\pi (k)}^{u}(t){v_{u,\pi (k)}}(t)\sqrt {P_{\pi (k)}^{u}(t)}\tag{12}\label{12}
 \vspace{-4mm}\end{equation}

\begin{equation}
{I_{\text {intra}}}_{\pi (k)}=\sum \limits _{i=k+1}^{K^{u}}x_{\pi (i)}^{u}(t)g_{\pi (i)}^{u}(t){v_{u,\pi (i)}}(t)\sqrt {P_{\pi (i)}^{u}(t)} \tag{13}\label{13}
\end{equation}

Building upon inter-cluster interference \eqref{8}, desired signal \eqref{12} and intra-cluster interference \eqref{13}, the SINR for $k$-th decoded user can be derived as \eqref{14} annotated at the bottom of previous page. Then \eqref{15} can be used to calculate the data rate of user $k$ connected to UAV $u$ with bandwidth $B$:

 \begin{equation*} \mathcal {R}_{\pi (k)}^{u}(t) = {B}\log 2\left ({{\gamma _{\pi (k)}^{u}(t)} + 1}\right),\tag{15}\label{15}\end{equation*}

\noindent{}The total data rate at time slot $t$ is the sum of data rate for all users associated with respective UAVs. This is written as 
\begin{equation*} 
{{\mathcal {R}(t)} = \sum \limits _{k = 1}^{K}\sum \limits _{u = 1}^{U}{{\mathcal {R}_{\pi (k)}^{u}(t)} } }.\tag{16}\label{16}
\vspace{-2mm}\end{equation*}

\noindent{}Therefore, the total system throughput for complete service time would be
\begin{equation*}
\vspace*{-2mm}\mathcal {R} = \sum _{t=0}^{T}\mathcal {R}(t)\tag{17}\label{17}
\vspace{-2mm}
\end{equation*}

\section{Problem formulation}
We aim to optimize the UAV 3D trajectory and power allocation policy for users in order to maximize the system throughput. However, the optimization is constrained by minimum QOS (data rate), maximum UAV transmission power and UAV spatial limits. This optimization problem can be expressed as \eqref{18a}. The positions of UAVs at any service time are expressed as $\mathbf {H} = \{0\leq t\leq T, 0\leq u\leq U, h_{u}(t)\}$. UAV velocity is assumed to be constant. $P_u$ denotes the UAV transmission power, $\mathbf {P} = \{p_{k}(t), 0\leq t\leq T, k \in {\mathbb K}\}$ is the power allocation policy and $\mathbf {V} = \{v_{k,u}(t), t = T_{r}, k\in {\mathbb K}, u\in {\mathbb U}\}$ represents the user-UAV serving indicator, which indicates user association. With that, the optimization function is defined as\vspace{-4mm}

\begin{align*}
\vspace{-2mm}
&\max _\textbf{P,V,H}
~\mathcal {R} = \sum _{t=0}^{T}\mathcal {R}(t), \tag{18a}\label{18a}\\&~\textrm {s.t.}~ x_{u}(t)\in\mathbb R:i\in[x_{\min},x_{\max}],\quad \forall t,~\forall u, \\&\hphantom {~\textrm {s.t.}~}y_{u}(t)\in\mathbb R:i\in[y_{\min},y_{\max}],\quad \forall t,~\forall u, \\&\hphantom {~\textrm {s.t.}~}h_{u}(t)\in\mathbb R:i\in[h_{\min},h_{\max}],\quad \forall t,~\forall u,\tag{18b}\label{18b}\\&\hphantom {~\textrm {s.t.}~}{L_{j}} \neq {L_{i}},\quad i,j ~\forall t,\in \mathbb {U},\tag{18c}\label{18c}\\&\hphantom {~\textrm {s.t.}~}{\sum _{u=1}^{N} v_{u,k}=1 }, \tag{18d}\label{18d}\\&\hphantom {~\textrm {s.t.}~}\sum \limits _{k \in {\mathbb K}} { {{v_{k,u}}(t)P_{u} \ge {P_{k}^{u}}} }, \quad \forall k,~\forall u,~\forall t, \tag{18e}\label{18e}\\&\hphantom {~\textrm {s.t.}~}G_{\pi (j)}^{u} \le G_{\pi (k)}^{u},\quad j < k,~\forall u,~\forall t,~\forall (j,k), \qquad \tag{18f}\label{18f}\\&\hphantom {~\textrm {s.t.}~}R_{k}(t) \geq R_{\text {QoS}},\quad \forall t,~\forall k, \tag{18g}\label{18g}\end{align*}
where \eqref{18b} to \eqref{18g} represent the constraints which the optimization is subject to. In that, \eqref{18b} indicates service area constraints imposed for possible 3D position of UAVs. \eqref{18c} ensures collision-free operations of UAVs when regarded as point particles. Constraint \eqref{18d} is introduced to ensure singular association of users with UAVs, i.e. each user is served by a single UAV. \eqref{18e} limits the maximum total transmission power of each UAV. \eqref{18f} indicates the decoding order for achieving successful SIC. \eqref{18g} details the user fairness constraint in terms of achieved datarate. It is pertinent to mention that the problem \eqref{18a} was demonstrated and categorized to be NP-hard in \cite{8624565}. Moreover, our scenario with mobile UAVs and users adds to the complexity to the problem and is therefore challenging to solve the formulated problem through the conventional convex-optimization algorithms. Therefore, the solution proposed involves the use of an RL-based algorithm, which has the ability to learn via its experience and interactions with the environment to find optimal solutions. 
\vspace{-1mm}
\section{Proposed Solution}
\label{secIV}
The proposed solution comprises of two steps: 1) a weighted K-means algorithm to determine optimum clusters and UAV associations for emergency vehicles; and 2) an SDQN-based method for joint optimization of UAV trajectory and power allocation policy. Weighted K-means clustering, described in first subsection, is adapted for spatial user clustering and association with upper bound limitation on cluster members to ensure adequate QoS. A multi-agent SDQN design is explained in the second subsection.

\vspace{-1mm}
\subsection{Weighted K-Means based Clustering} K-means algorithm has promoted favorable performance with respect to user clustering in wireless communications \cite{javed2023uav}. The low computation complexity of K-means algorithm enables timely clustering of all users and arrests service interruptions when performing periodic re-clustering. Another reason for choosing K-means algorithm is that it does not require a history of user movement and only relies on the present state to establish optimum clusters. However, the standard K-means approach will fail to take the current position of UAVs into account while clustering. To incorporate UAV positions, we invoke weighted K-means clustering with higher weights allocated to the UAV locations. 

Let user positions be denoted by $\mathbb {L}_{User}=\{ l_{1}, l_{2}\ldots l_{K}\}$ and 3D UAV positions by $\mathbb {L}_{UAV}=\{ L_{1}, L_{2}\ldots L_{u}\}$. Then $\mathbb {L} = \{ \mathbb {L}_{User}, \mathbb {L}_{UAV}\}$ and weights $\mathit {\Phi} = \{\phi_{User}, \phi_{UAV}\}$ serve as the input for calculating clusters such that   $\phi_{UAV} = 2 \cdot  \phi_{User}$. Clusters can be calculated on any node and communicated amongst UAVs through a control channel. $U$ clusters $(C_{1}, C_{2} \ldots C_{U})$ are then calculated such that all cluster members are closely located to each other. The algorithm first arbitrarily chooses $U$ positions as cluster centers and then assigns each member to the closest cluster center. The centroid of each cluster is recalculated to minimize sum of squared error (SSE) using \eqref{19}. These steps are re-iterated until the centroids of all clusters do not change.

\begin{equation*} SSE = \sum _{i=1}^{U}\sum _{l\in C_{i}} \| l - \mu _{i}\|^{2},\tag{19}\label{19}\end{equation*}
where $\mu _{i}$, known as centroid, is the weighted vector mean of cluster $C_i$ and calculated as 
\begin{equation*} \mu _{i} = \frac {1}{|C_{i}|} \sum _{l\in C_{i}} l \cdot \phi.\tag{20}\label{20}\end{equation*}

It is worth mentioning that UAVs have a maximum number of users they can serve. This constraint, however, is not taken into account using the above-mentioned clustering algorithm. Therefore, an add-on scheme is required in our case to limit the number of users in any cluster. We do this by identifying clusters with an excess number of users. The farthest user in that cluster is re-associating with the second-nearest cluster.
This method is iterated until the number of users associated with each cluster is within the UAV's service capacity. \textbf{~Algorithm~\ref{alg:one}} delineates steps involved in user clustering.

\begin{algorithm}[hbt!]
\scriptsize
\SetAlgoLined
\caption{User Clustering Algorithm}\label{alg:one}
Initialize number of clusters $|(C1\ldots C_U)$, maximum users allowed in a cluster $\upsilon$, maximum iterations $N$\\
Get users location $\mathbb {L}_{User}=\{ l_{1}, l_{2}\ldots l_{K}\}$ and UAV locations $\mathbb {L}_{UAV}=\{ L_{1}, L_{2}\ldots L_{u}\}$\\
Input $\mathbb {L}$ = \{ $\mathbb {L}_{User}$, $\mathbb {L}_{UAV}\}$\\
Input weights $\mathit {\Phi}$ = \{$\phi_{User}, \phi_{UAV}$\}\\

Randomly choose $U$ co-ordinates as initial centroids $(\mu _1 \ldots \mu_U)$ from $\mathbb{L}$\\

\While {$n \leq N$}
    {\For{$l \in \mathbb{L}_{User}$}
        {Calculate $d_{uk} = \|l_k - \mu_u\|$\\
        For $C_u$ with least $d_{uk}$, assign $l_k$ \\
        Recalculate $\mu_u$ as per \eqref{20} }
     \If{$\mu_u(n-1) = \mu_u(n)$}
        {Exit Loop}
    }

\While{$|C_{u}| > \upsilon$}
    {Unassign furthest $l_k$ from $C_u$  \\
    For $C_i$ with least $d_{ik}$ s.t. $i \neq u$ , assign $l_k$}
Output $|(C_1\ldots C_U)$
\end{algorithm}
\vspace{-3mm}

\subsection{Trajectory and Power Allocation using MDQN}
A multi-agent SDQN algorithm can be used to train numerous UAVs via the same DQN to achieve joint optimization of 3D trajectory of UAVs and user power allocation. Each UAV acts as an independent agent and is free to choose actions. Although UAVs are not aware of each other's actions, they can all connect to the same DQN through state abstraction. State abstraction is the re-organization of experiences of each UAV in a standard form so that it can be input to the same NN for shared learning. This helps catalyse the learning process as NN is provided multiple experiences in a single timestep, i.e., training of each agent is bolstered by experiences shared by other UAVs in the fleet. Consequently, the training time is reduced and the problem of prolonged training periods of a conventional DQN is alleviated. 

It should be noted that SDQN requires data exchange only for training, during which the updated NN parameters are then copied to each agent. However during service, the agent may choose either to use previously stored NN parameters or update its NN parameters before making predictions. Therefore, no additional communication is required between agents with an SDQN paradigm. Additionally, closely located UAVs can increase inter-cluster interference and, consequently, decrease the data rate and DQN reward. This encourages UAVs to intelligently stay away from one another and avoid interference, thus, in a way, achieving indirect cooperation. This is made possible by considering UAVs as part of the environment and including positions of other UAVs as state information while making decisions. 

\subsubsection{Conventional and Deep Q-Learning}

DQN is a value-based, model-free reinforcement learning algorithm. The agent chooses from a number of actions in a dynamic environment such that it maximizes its long-term reward. The action value, also known as Q-value, indicates the cumulative discounted reward for each action ($A$). During each time slot, the agent observes a state ($S$) and has to choose actions from the action space. In our scenario, state is defined by the UAV fleet positions and channel state information (CSI) of users. The reward ($R$) denotes the quality of action taken by a UAV and is dictated by two things:  the sum rate of users served by that UAV, and whether all users meet a pre-defined QoS criteria. A table is introduced in Q-learning algorithm which contains Q-values of all actions for each state. The agents can then choose actions with the maximum Q-value in a given state. However, complex systems with large states and actions spaces require huge memory to constantly update and store the Q-table. Therefore, for such cases, it was proposed that Q-table is replaced by a deep NN \cite{mnih2015human}, which estimates the Q-value of each action. This enables the agent to update the NN through training, and choose actions as per the Q-value estimated by the DQN. DQN can also mitigate the Q-learning algorithm's generality issue by predicting Q-values even for unseen states through its neural network~\cite{iosifidis2022deep}.

During each time step, the agent observes the current state $S$, then selects an action $A$ as per the action policy. The action, comprising of a set of flight maneuvers and power allocation policy for users, changes the environment and, therefore, the state from $S$ to $S'$ following a Markov Process \cite{9044134}. $S$, $A$, $R$ and $S'$ are saved by the agent as an experience from that time step to train the NN. The DQN always chooses action with maximum Q value to maximize the long-term reward. The Q-value is calculated through the action-value function, described by the Bellman equation
\vspace{-4mm}

\begin{align*} Q(S,A) \leftarrow Q(S,A) + \alpha [R + \beta \max Q(S',A') - Q(S,A)]. \\\tag{21}\end{align*}
\vspace{-4mm}

\noindent{}where $\alpha$ ($0 \leq \alpha \leq 1$) denotes the learning rate and $\beta$ ($0 \leq \beta \leq 1$) represents the discount factor. 

However, in cases with large state and action spaces such as ours, the action-value function is approximated by a neural network. In this regard, numerous studies have corroborated the suitability to fit complex functional relations. However, training each agent separately for the same task would be slow and computationally taxing. Using an SDQN allows agents to share experience and to train the same NN, hence decreasing the training time and increasing the computational efficiency of the system.

\subsubsection{State Abstraction for Input} All DQN models need state information as input to estimate Q-values for actions. In this case, state information comprises 3D position of UAVs and user channel information of users. The advantage of using user channel information, instead of user location, is that channel information is inherently known to the UAVs whereas the collection of user location each time causes communication overheads. Therefore, user channel information also forms the basis for deciding power allocation for all users. 

\begin{figure*}[!t]
\vspace*{-4mm}
\centering
\subfloat[]{\includegraphics[width=0.9\columnwidth]{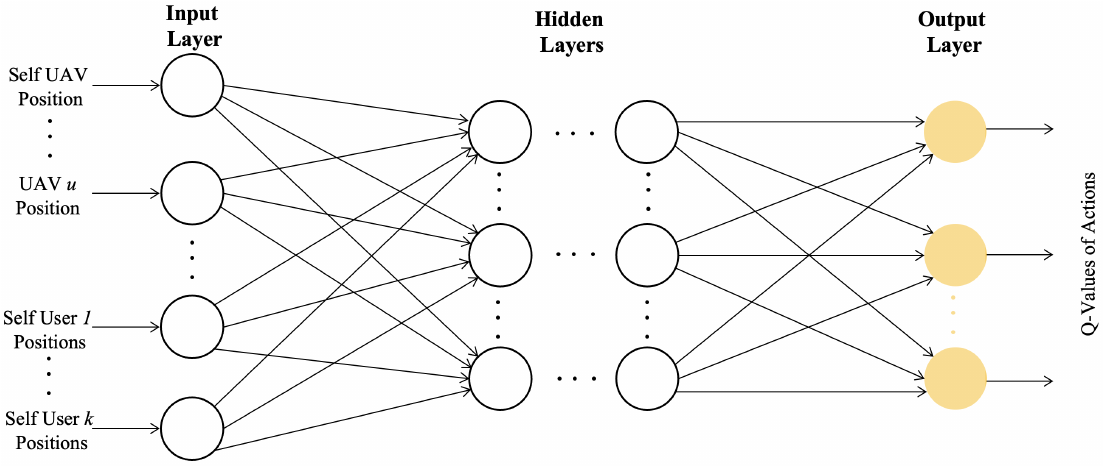}%
\label{conventional_dqn_architecture}}
\hfil
\subfloat[]{\includegraphics[width=0.9\columnwidth]{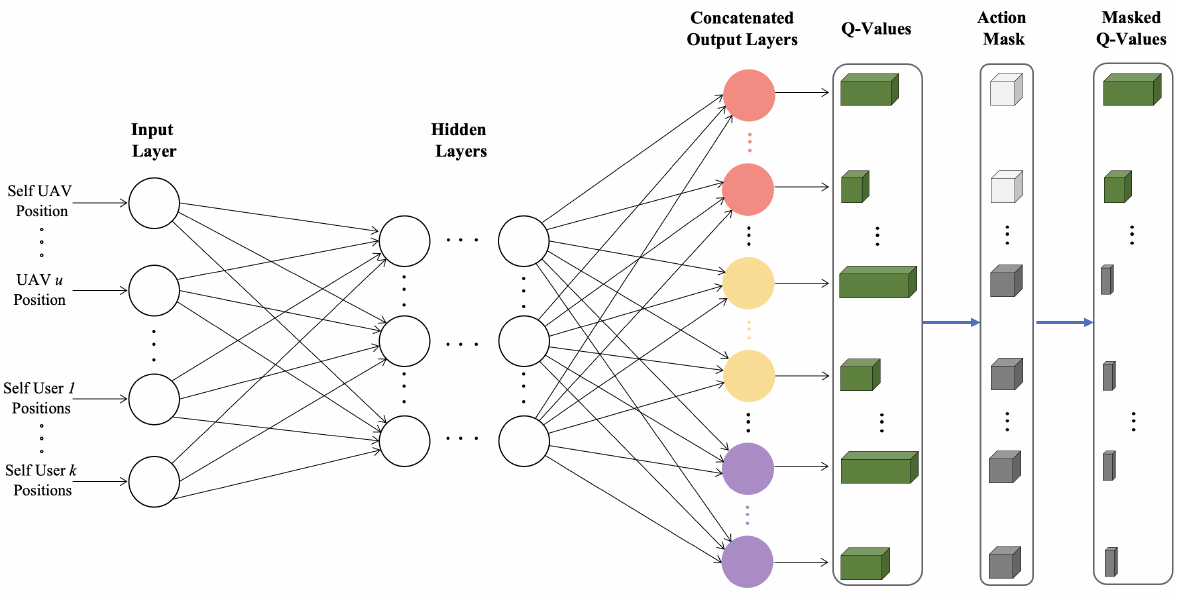}%
\label{sdqn_architecture}}
\caption{Comparison of NN input connections and output nodes between (a) DQN and (b) SDQN with action masking}
\label{architecture_comparison}
\vspace{-3mm}
\end{figure*}

However, all agents share the same NN in an SDQN algorithm. Therefore, it is binding for all UAVs to rearrange their state information into a standard form before so that it can be fed to the SDQN input layer. We invoke the reshuffling method proposed in \cite{9520121}. The rearrangement of state information into a standard array is depicted in the input layer of Fig.~\ref{architecture_comparison}. To explain, when UAV1 is connected to the NN, its location information is fed into the first neuron. The following input nodes are fed the location information of all other UAVs in the fleet. The subsequent nodes are fed channel gains of users served UAV1, followed by channel gains of all other users. Similarly, when UAV2 is connected, the first node carries UAV2 location information. By this scheme, each UAV feeds its location and served users' channel information into specific input nodes. This scheme allows the neural network to draw a global relationship between each UAV and its users. The state is expressed as 
\begin{equation} 
S = \{L_{u}(t),L_{s}(t),g_{k}^{u}(t),g_{k}^{s}(t)\}, \quad u,j \in \mathbb {U}, ~j\neq u, ~k \in \mathbb {K},\!\! \\\tag{22}\end{equation}
where $L_u(t)$ represents 3D coordinates of agent currently connected with NN and $L_s(t)$ denotes the co-ordinates of all other agents, which serve as source for inter-cluster interference. Similarly, $g_k^u(t)$ are the channel gains of users served by UAV connected to the NN and other users, whereas $g_k^u(t)$ represents channel gain of other users.

Due to the fact that number values of $g_k^u$ and $L_u$ differ in the magnitudes of high order, scaling is done while pre-processing the data for fast network convergence.

\noindent{\textit{Remark 2:}} \emph{As opposed to the use of different state space for each agent and training them separately, state abstraction allows numerous agents to train a neural network jointly. This, in turn, provides the NN with more episodes to train on in a time period and hence catalyzes the training process. State abstraction, which includes re-ordering of data for input and scalarization, is deemed vital to ensure that NN converges.}

\subsubsection{Invalid Action Masking}
\label{invalid action masking}
In contrast to DQN, in which agents train their models independently, an SDQN scheme allows all agents to connect sequentially and train the NN at each time step. State abstraction allows each agent to re-arrange the observed state and provide it as common input to the NN. However, number of possible outputs for each agent can vary depending upon the number of users served. This is the case because transmission power needs to be distributed amongst the served users. For example, NOMA power allocation policy for 2 users will distribute power into 2 proportions, whereas the policy for 4 users will be distributed into 4 proportions. This translates into a variable number of output nodes for power allocation policy (2 and 4, respectively, in the previous example). The number of output nodes has a positive co-relation with the number of users being served by any UAV.

We circumvent this dilemma of shared learning of a NN with agents consisting of varying action spaces size by concatenating all possible action spaces. However, concatenating allows a UAV serving 2 users to choose actions of a UAV serving 4 users. This is counter intuitive and deleterious to the learning process, as choosing `impossible actions' is meaningless and wasteful. We use action masking \cite{huang2020closer} to ensure that an agent does not choose impossible actions. Dictated by the number of users being served by an agent, we invoke invalid action masking - whereby the Q-value of impossible actions is replaced by -$\infty$. The application of action mask thwarts the possibility of Q-value of invalid actions to be maximum, thereby reducing the probability of their selection to zero. Action space concatenation and use of invalid action masking is depicted in Fig. \ref{architecture_comparison}.
\begin{figure*}
    \centering
  \includegraphics[scale=0.7]{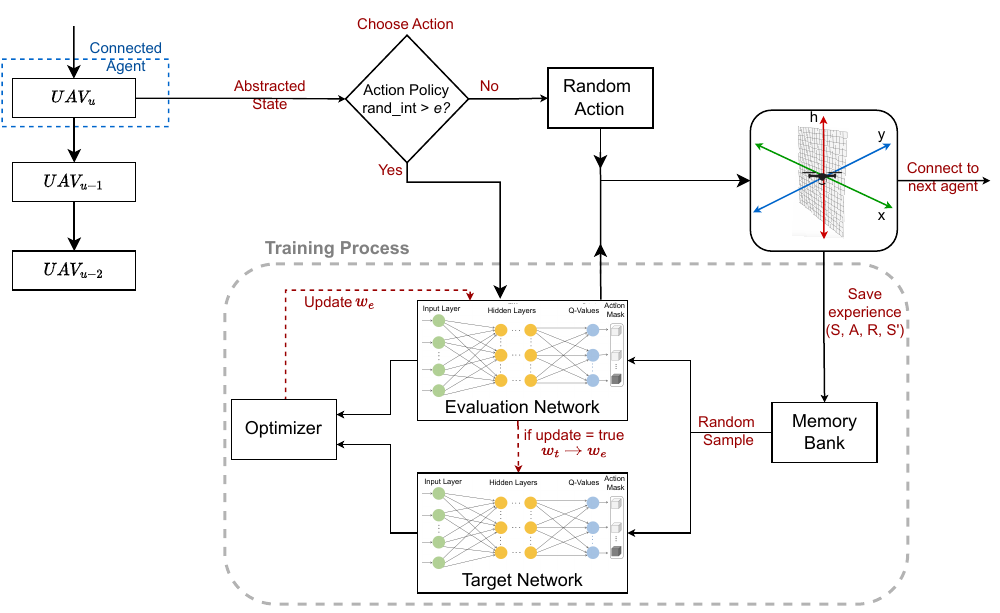}
  \caption{Schematic diagram of SDQN based Trajectory design and Power allocation}
  \label{SDQN_schematic}
\vspace{-4mm}
\end{figure*}

\subsubsection{SDQN Algorithm}
Fig. \ref{SDQN_schematic} provides a schematic diagram of the SDQN algorithm. Each UAV sequentially inputs the abstracted state into an evaluation network to determine optimal action and extract rewards. After respective actions have been completed, the sum rate of all users is calculated and added to determine the system data rate for that time slot. 

The SDQN algorithm introduces a target network with the same architecture as that of the evaluation network. During training, the target network is tasked to estimate target Q-values, which are juxtaposed against estimated Q-values provided by the evaluation network to compute loss. The reason for not using the same network to estimate both values is that during each training step, the values of the evaluation network vary. If used to make corrections, the shifting target values can destabilize the network by vacillating between target and estimated Q-values. Therefore, the target network parameters are not changed with every step but rather updated to the latest evaluation network parameters periodically. The MDQN algorithm flow is given in \textbf{Algorithm~\ref{alg:two}}.

\begin{algorithm}[hbt!]
\scriptsize

\caption{SDQN for Trajectory and Power allocation}\label{alg:two}
\SetInd{0.2em}{0.8em}

\For{each episode}
    {Initialize UAVs and users' positions\\
    Initialize random parameters for evaluation network $w_e$ and target network $w_t$\\
    Initialize experience replay memory $M$\\
    Update $\epsilon$ with $\epsilon$-decay
    
    \For{every time step $t \in\mathbb Z:i\in[t_0,t_0+T_r]$ }
        {\For{all UAVs}
            {Calculate $G_k^u$, k $\in \mathbb{L}$ \\
            Create abstracted state array $S$  \\
            Observe number of users in cluster, $|C_u|$ \\
            Based on $|C_u|$, generate and apply action mask \\
            Choose masked action $\Bar{A}$ as per action policy \\
            Execute $\Bar{A}$, observe Reward $R$ and next state $S'$ \\
            Store experience $e = (S,\Bar{A},R,S')$ in $M$ \\
           
           \If{enough experiences in $M$} 
                {Sample $minibatch \omega$ from transitions in $M$ 
                
                \For{each $(S_i,\Bar{A}_i,R_i,S_i')$ in $\omega$}
                    {
                    $y_i = R_i + \beta \max Q(S_i',\Bar{A}_i',w_{t})$\\ 
                    Calculate Loss $\mathcal{L} =1/\omega \sum _{i=0}^{\omega-1}(Q(S_i,\Bar{A}_i,w_e)-y_i)^2$
                    Update $w_e$ using the SGD algorithm\\
                    Copy $w_e$ to $w_t$ every $v$ steps\\
                    }
                }
            $S \leftarrow S'$ \\
            } 
        Move users
        }
    }

\end{algorithm}

\subsubsection{Action Space} Action space size of UAVs can differ as they can serve clusters with different numbers of associated users. Action space for SDQN is generated by concatenating outputs for all possible cluster sizes. Each output node dictates UAV to take two actions, choose power allocation policy, and decide the UAV direction. Discrete action space is used since a continuous action space would translate into infinite actions to choose from, resulting in significant memory overheads. Therefore, agents are required to choose a preset action for movement and power allocation policy.

\textbf{Movement action:} Agents are allowed to choose between seven movement directions, namely: up, down, forward, backward, left, right, and hover. If a chosen action leads a UAV to fly out of the service area, the hover action is chosen as default.

\textbf{Power allocation policy:} Depending on the number of users being served, action space for distribution of transmitted power for users is pre-fixed in multiple power gears, and is denoted by $p_1, p_2,\ldots,p_x$. The selected power allocation is sustained until the next action is invoked. 

\begin{figure}
    \centering
  \includegraphics[scale=0.5]{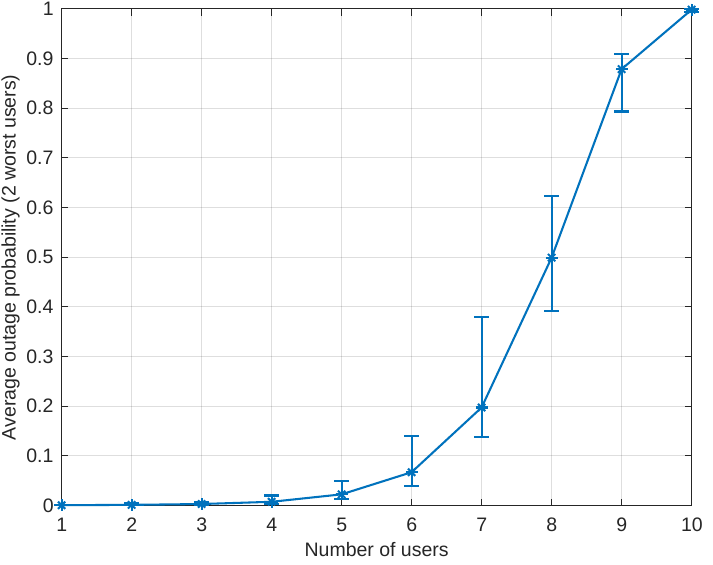}
  \caption{Outage Probability simulation of worst users with parameters in Table~\ref{Simulation parameters}. Mean outage of worst users increases as the number of single-channel NOMA users increases. The error bars depict the maximum and minimum mean outages in 1000 Monte-Carlo simulations.}
  \label{outage_prob_SC}
\vspace{-4mm}
\end{figure}

\textbf{Scalability:} The system is scaled by varying the number of UAVs or users. As the SDQN architecture exploits experience sharing, increasing the number of UAVs will result in efficient learning. However, catering to a very large number of users in the single-carrier NOMA system would result in a bigger action space and may cause uncertainty regarding convergence. Even Deep Deterministic Policy Gradients, which are known to operate over continuous action dimensions, would struggle in this scenario. This is because the power allocated to the strongest user decreases exponentially as the number of NOMA users increases and causes a decreased achievable data rate and frequent outages. Fig. \ref{outage_prob_SC} (simulated using parameters in \textbf{Table~\ref{Simulation parameters}}) shows that as the number of NOMA users in a single carrier system increases, the mean outage of the worst users increases and makes it increasingly hard to overcome the QoS constraint  in \eqref{18g}. Moreover, a large cluster size may also cause degraded NOMA performance due to error propagation in SIC.  Therefore, the preferred method to cater to a large number of users is to increase the number of service UAVs. One alternative, left for future work, may be to incorporate hybrid NOMA schemes by using multiple carrier channels and distributing users overall. 

\subsubsection{DQN Training} The neural network is required to be trained with a sufficient number of instances to accurately approximate the Q-value. However, during training, consecutive samples are not selected as it could translate into high correlation in the training dataset. This issue is bypassed by using a replay memory technique. This technique allows the agent to take actions at random and store the experience ($S, R, \Bar{A}, S'$) in the replay memory until it is full. It is only after this that the training starts by choosing experiences at random from the memory bank. These experiences have significant time differences and, hence, reduced sample correlation. The DQN trains parameter $w$ such that it minimizes the loss function $J()$ per \eqref{24}, where $y$ represents the target and $J()$ can be tailored to suit as per the dictates of the optimization problem. 
\vspace{-2mm}

\begin{equation*} 
\vspace{-3mm}
y=R + \beta \max Q(S',\Bar{A}') \tag{23}\label{23}\end{equation*}

\begin{equation*}
J(w)=(y - Q(S,\Bar{A},w))^{2}\tag{24}\label{24}\end{equation*}

As training progresses, the agent probabilistically decides whether to exploit prior knowledge and take optimal action as per action policy or explore for better reward by taking random action. 

\subsubsection{Action Policy} The epsilon-greedy ($\epsilon-greedy$) policy, with a decreasing value for $\epsilon$, is chosen to deliver a balance between exploratory actions earlier in the training, and exploitation actions as the training progresses. With this policy, the UAVs explore the environment by taking random action with probability $\epsilon$ and exploiting available information with probability $\epsilon-1$.
\vspace{-4mm}

\begin{align*} 
\Bar{A}_{(t)}= \begin{cases} \displaystyle any~action, & \epsilon, \\ \displaystyle {argmax}_{\Bar{A}} Q(S,\Bar{A},w_{e}),& 1-\epsilon. \end{cases}\tag{25}\end{align*}

\subsubsection{Reward} The objective function, as defined in \eqref{18a}, aims to maximize the total throughput of the multi-UAV system while adhering to constraints as dictated in \eqref{18b} to \eqref{18f}. For a multi-UAV system, it is imperative to base the reward on sum of data rate of all UAVs, rather than UAVs' own rate. Such an arrangement would allow UAVs to choose actions that benefit the throughput of system as a whole. These decisions will be based on the the State information. Moreover, to emphasize importance of meeting the QoS requirement in \eqref{18g}, a penalty co-efficient $\lambda$ is introduced in the reward function. Set 0 by default, $\lambda$ increases if a UAV chooses an action that denies QoS. However, $\lambda$ would not rise in situations where taking any action would not satisfy the QoS requirements. The reward function, then, is given by

\begin{equation*} R_{u} =\frac {\mathcal {R}(t)}{2^\lambda },\quad u \in U,\tag{26}\end{equation*}
where $\mathcal {R}(t)$ is the sum data rate and $\lambda$ is the penalty co-efficient.

\begin{table}[h]
\caption{Simulation Parameters.}
\centering
\begin{tabular}{|c| c| c| } 
    \hline
    Parameter & Notation & Quantity \\
    \hline
    Carrier frequency& $f_c$ & 2 GHz \\  
    \hline
    UAV fleet size& $U$ & 3 \\  
    \hline
    Number of users & $K$ & 6 \\  
    \hline
    Channel bandwidth & B & 15 kHz \\  
    \hline
    Maximum user speed& $V_{max}$ & 10 m/s \\  
    \hline
    UAV speed & $V$ & 5 m/s \\  
    \hline
    Transmission power& $P_{max}$ & 29 dBm \\  
    \hline
    Dynamic re-clustering interval & $T_r$ & 60 s \\  
    \hline
    UAV height & $h$ & [20 m, 150 m]  \\  
    \hline
    Minimum service area & $x_{min}, y_{min}$ & 0 m \\  
    \hline
    Maximum service area & $x_{max}, y_{max}$ & 1000 m  \\  
    \hline
    QoS requirement & $R_{QoS}$ & 0.15 kb/s \\  
    \hline
    AWGN power & $\sigma$ & -100 dBm/Hz \\  
    \hline
    Discount Factor & $\beta$ & 1 \\  
    \hline
     Replay memory size& $e$ & 10000  \\  
    \hline
    Batch size & $\omega$ & 128 \\  
    \hline
    Minimum learning rate & $\alpha$ & 0.001 \\  
    \hline
     Greedy co-efficient& $\epsilon$ & $\leq0.9$ \\  
    \hline
     Target DQN update frequency& $v$ & 1000-2000 \\  
    \hline

    \end{tabular}
\label{Simulation parameters}
\vspace{-6mm}
\end{table}

\subsection{Complexity of Algorithms}
\subsubsection{Clustering} Complexity of the weighted K-means clustering algorithm to cluster $K$ users in $U$ clusters with N maximum iterations is given as $ O(K \cdot U \cdot N)$. Cost of correct user association with UAVs is $ O(N \cdot U^2)$. Since $U << N$, the overall complexity can be written as $ O(K \cdot U \cdot N)$.
\subsubsection{Selecting an Action} Assuming an MDQN architecture with $|S_n|$ number of input nodes, $h$ fully connected hidden layers with $\zeta_i$ nodes in layer $i$ and $|A_u|$ possible actions UAV $u$. Since there are separate action spaces for UAVs depending upon the number of users served, the complexity for selecting action for UAV $u$ is $ O(|S_n| \cdot \zeta_1 + \sum_{l=1}^{h} \zeta_l \cdot \zeta_{l+1} + \zeta_{h} \cdot |A_u|)$ 
and is denoted $O(C_u^a)$. Note that due to state abstraction $|S_n|= K + 3U$ and $|A_u| = p_x \cdot K_u + 7U$ where $K_u$ represents number of users served by UAV $u$ and $p_x$ represents the number of pre-fixed power gears as discussed in power allocation policy.

\subsubsection{Training Complexity} Complexity for back propagation algorithm includes gradient computations and weight updates and is similar to that of forward pass. Therefore, the overall training complexity with $E$ episodes, $B_u$ batch size for each UAV and $t$ steps is approximated as $O(E \cdot T \cdot  \sum_{u=1}^{U}(B_u \cdot C_u^a))$. For a 3 layer neural network with $n$ hidden nodes, this simplifies to $O(E \cdot T \cdot |S_n| \cdot n \sum_{u=1}^{U}(B_u \cdot |A_u|))$
\vspace{-1mm}

\begin{figure*}[!b]
\vspace*{-2mm}
\minipage{0.37\textwidth}
\centering
    \includegraphics[scale = 0.36]{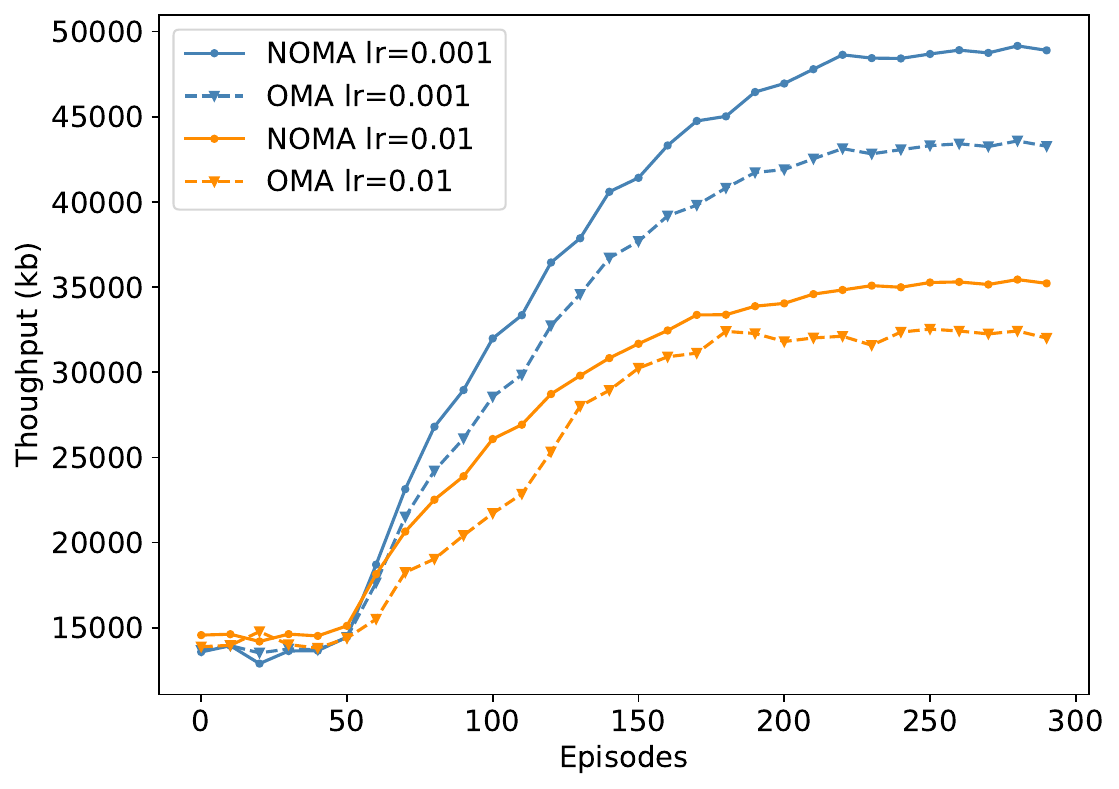}
    \caption{Training comparison for NOMA vs OMA schemes with respect to throughput.}
    \label{NOMA_vs_OMA_lr}
\endminipage\hfill
\minipage{0.6\textwidth}
  \subfloat[]{\includegraphics[scale = 0.298]{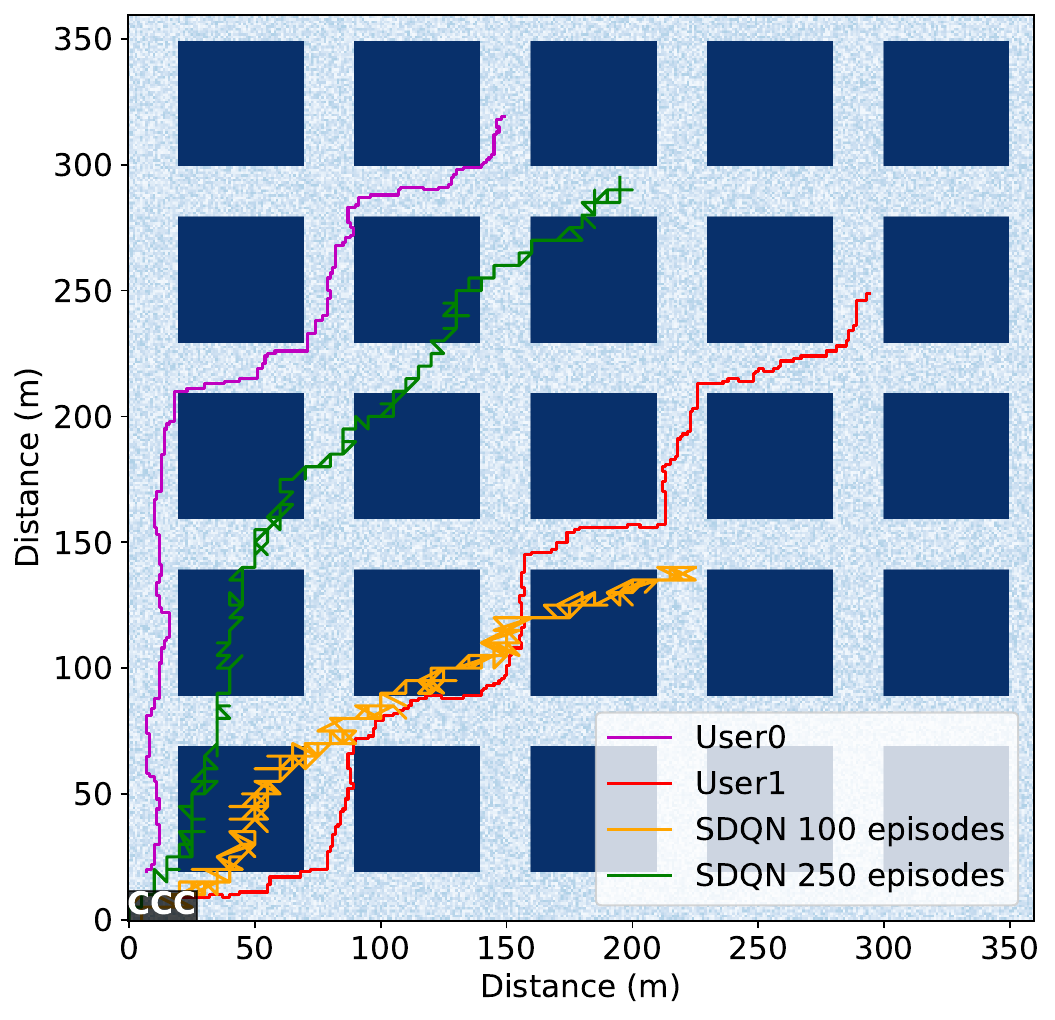}%
\vspace{-1mm}
\label{UAV trajectory plan view}}
\hfil
\subfloat[]{\includegraphics[scale = 0.298]{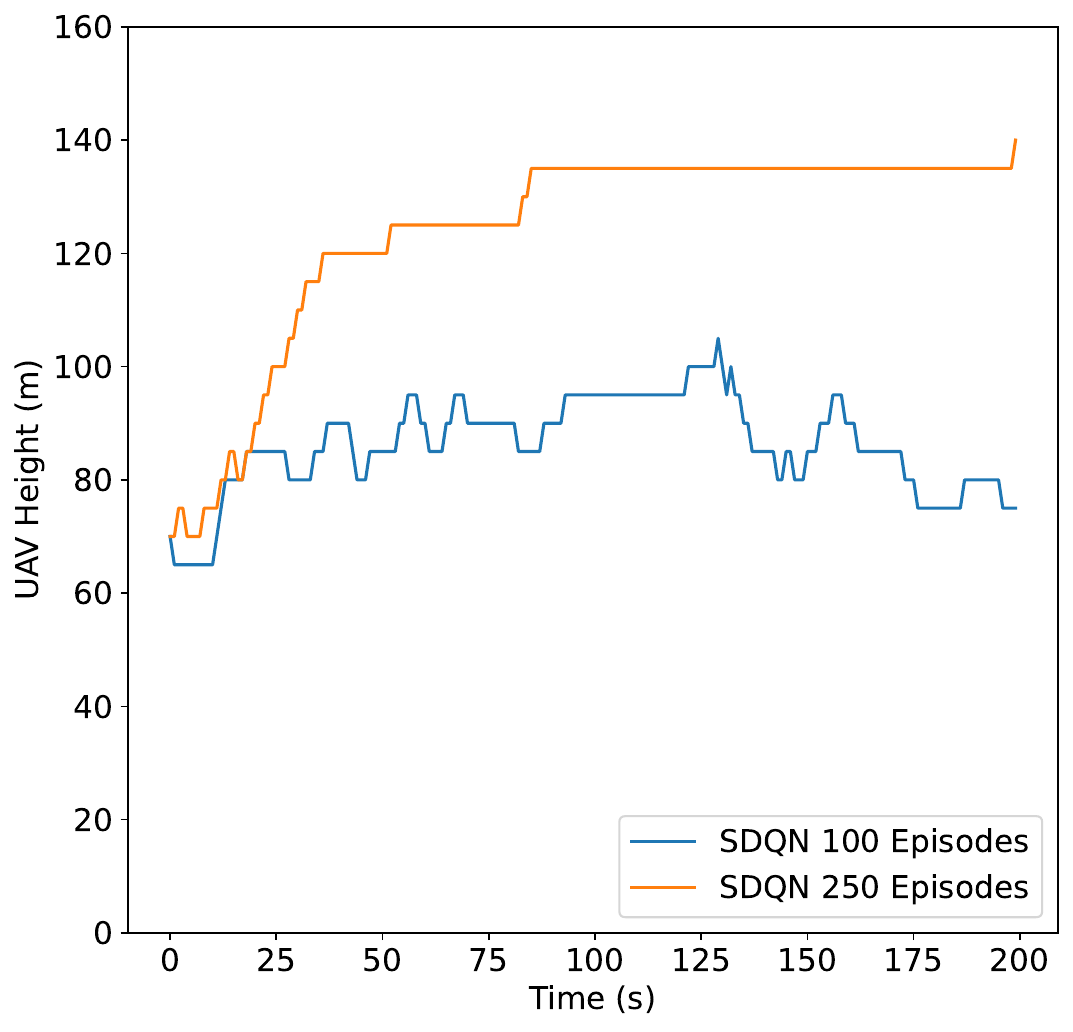}%
\vspace{-1mm}
\label{Height Profile}}
\caption{Comparison of 3D trajectory using SDQN after 100 and 250 training episode.}
\label{2_uav_2_users_trajectory}
\endminipage
\vspace{-4mm}
\end{figure*}

\section{Simulation Results} 

We perform simulations on an Apple M1 Pro with 3.2 GHz CPU and 16 GB RAM. The operating system is macOS Ventura 13.2.1. We train the DQN on an M1 Pro GPU with 16 cores and program with TensorFlow 2.5 in the JupyterLab environment. The conventional DQN algorithm, circular UAV trajectories, and mutual learning per \cite{9520121} are chosen as benchmarks against which to examine performance improvements. In the simulation, the users start of from the CCC (origin) with $V_{max}$ = 10 m/s, which follows from \cite{eze2010heterogeneous}, and follow paths to destinations dictated by Dijkstra's Algorithm. UAVs start randomly near the mean allowable altitude to avoid extreme initialization. NN has 3 layers with a hidden layer consisting of 70 nodes. Rectified linear units and mean squared error are selected as activation and loss functions, respectively. The Adam optimizer is used to train the Network. Greedy action policy is used with $\epsilon$ linearly decreasing from 0.9. Other default simulation parameters are tabulated in \textbf{Table~\ref{Simulation parameters}}. Unless stated otherwise, simulations have been carried out with default parameters.

Fig. \ref{NOMA_vs_OMA_lr} is a throughput versus training episodes plot comparing SDQN algorithm for both NOMA and OMA. The graph demonstrates that NOMA-assisted SDQN performs better in terms of convergence and final throughput yield. It can be noticed that there was almost no throughput increase in the initial 50 episodes for any case. This can be attributed to two reasons: 1) the replay memory is being filled with training experiences from all agents and training does not start until its completely filled; and 2) large $\epsilon$, which translates into large probability of random action and encourages the system to take exploratory actions. 

It is worth noting that when the rate is relatively high, i.e. 0.01, training becomes comparably less stable and the improvement in throughput is not as pronounced. However, NOMA scheme performs better than its OMA counterpart for both rates. With the learning rate 0.001, NOMA yielded 13\% higher throughput compared to the OMA technique.


Fig. \ref{2_uav_2_users_trajectory} exhibits the 2D trajectory and altitude profile of a UAV serving a NOMA cluster as the users move outward from near the origin (simulated as CCC). SDQN model weights were saved and tested after 100 and 250 training episodes to see difference in trajectory between partially trained and well-trained model. It is evident that SDQN model tested after 100 episodes juggles between positions and struggles to find the right 3D location. This is because the model has been trained for only a few episodes after repleting experience replay memory. $\epsilon$ is still high and so still prefers exploring as compared to capitalizing on and refining already held Q-values. The UAV seems to favor an overall increase in amplitude but fails to establish any consistency in height or 2D coordinate, and therefore wanders off in a wayward direction as time progresses. On the other hand, after 250 episodes SDQN model is almost fully trained and follows a better trajectory. As the distance between users increases, the UAV quickly gains height to about 125 m and maintains a similar height throughout as it follows the users in the 2D plane.

\begin{figure*}[!htb]
\vspace*{-4mm}
\minipage{0.325\textwidth}
  \includegraphics[width=\textwidth]{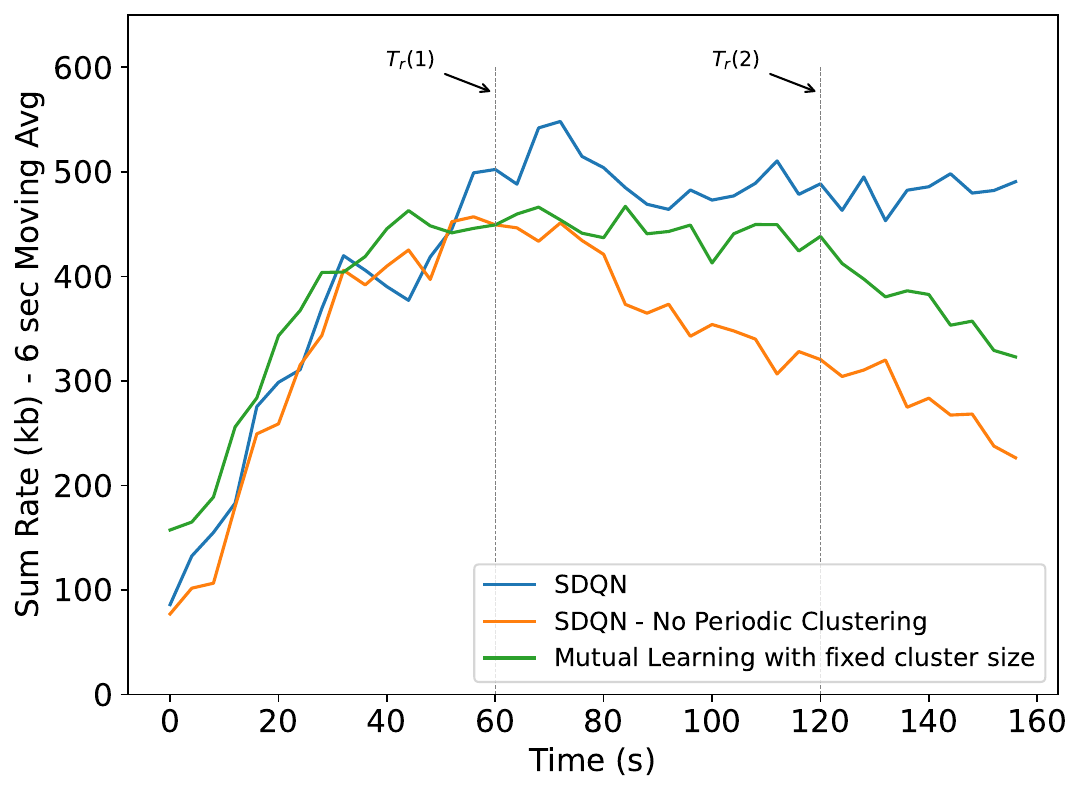}
     \caption{System data rate comparison for SDQN model with/without clustering and Mutual learning model with fixed cluster size.}\label{reclustering}
\endminipage\hfill
\minipage{0.325\textwidth}
  \includegraphics[width=\textwidth]{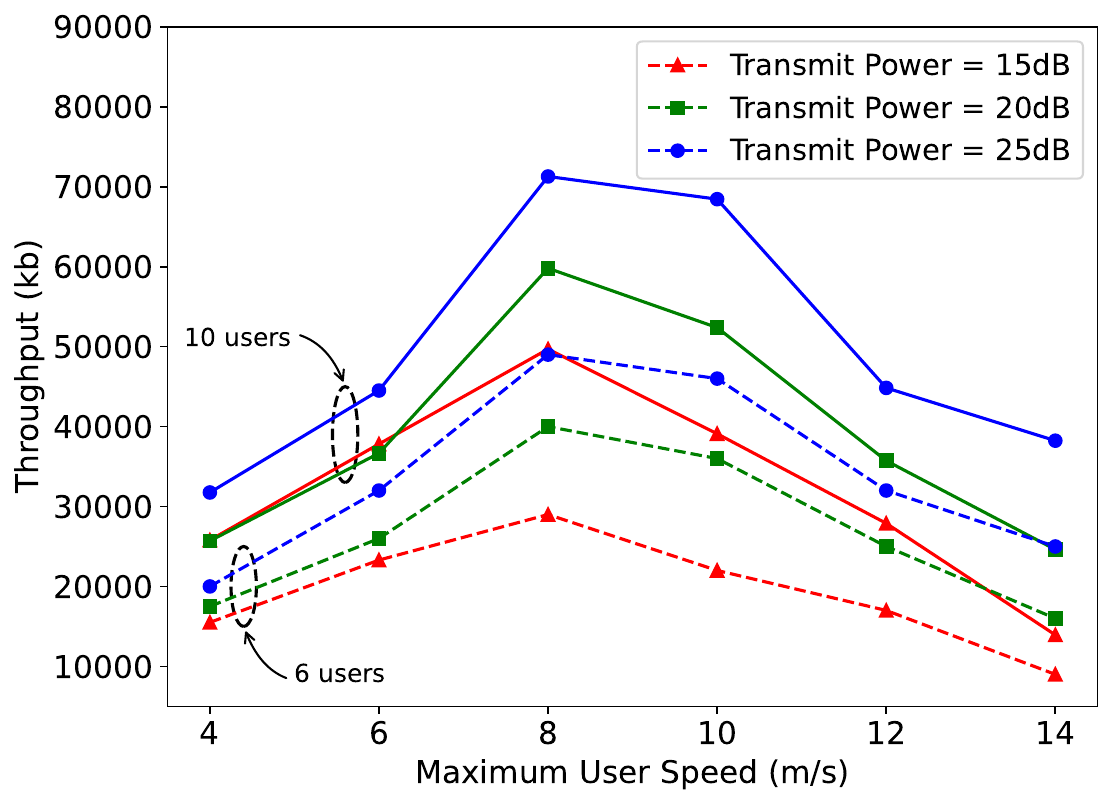}
     \caption{Throughput comparison for SDQN model for different user speeds and maximum UAV transmission power.}\label{Throughput_vs_Maximum_User_Speed}
\endminipage\hfill
\minipage{0.325\textwidth}%
  \includegraphics[width=\textwidth]{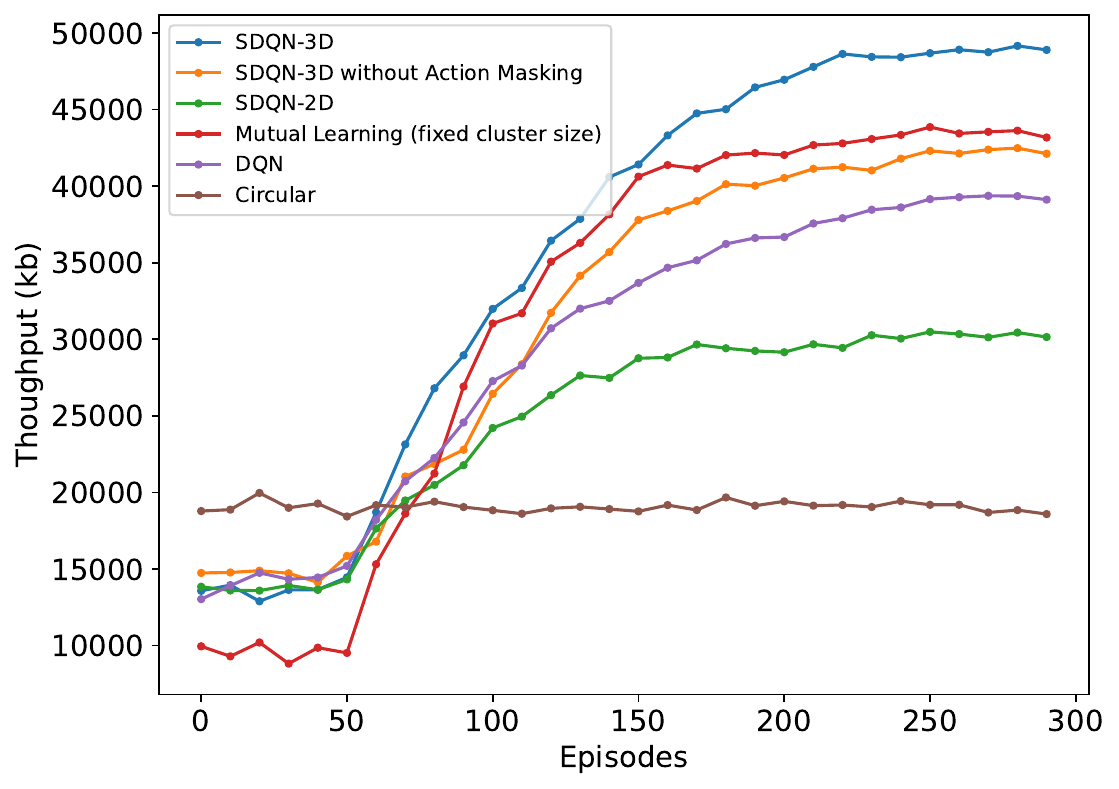}
     \caption{Throughput improvements using the proposed solution over other trajectories and training schemes.}\label{Throughput_vs_trajectory_comparison}
\endminipage
\vspace{-4mm}
\end{figure*}

Fig. \ref{reclustering} helps draw a comparison between data rates of SDQN scheme with and without periodic re-clustering, and Mutual learning scheme with the same action space for all users used in \cite{9520121}. To check the scalability of the proposed solution, 6 users start off close to the CCC and employ three UAVs. For the two SDQN models, same parameters were used except that one model does not perform periodic clustering. Selecting a moderate $T_r$ (60s) allows the users to move enough distance to bring about a significant change in the UAV-user channel gains and make re-clustering meaningful. During the first 60 secs, both SDQN cases have 2 users each being served by all UAVs and therefore the performance is similar. UAVs find their optimum location based on current clusters quickly and hence the data rate increases sharply. However, after $T_r(1)$, the first clustering interval, the users stray away from each other towards their respective sparsely placed destinations and the initial clustering loses relevance. At this point, the proposed model re-clusters users and assigns them to the nearest UAV. In this way, the proposed model constantly revises clusters to optimize user-UAV association. On the contrary, data rate for non-reclustered case starts declining after being relatively stationary between 80 and 110 sec. As the users in a cluster continue to be served by the same UAV, however orthogonal their movements, the difference in data rate between the two schemes balloons even more after the second clustering interval. 

The mutual learning algorithm proposed by \cite{9520121} allocates the same cluster size to all agents so that the same action space can be used for learning. For a fair comparison, this model uses periodic re-clustering and all agents are trained for 2 user clusters exclusively. For the initial clustering, the mutual learning algorithm performs 10\% better than the SDQN model, which is optimized to serve 1 to 3 users in a cluster. However, as the users are re-clustered at $T_r(1)$, the K-means algorithm yields different-sized clusters. Here the mutual learning scheme forcibly allocates 2 users to each cluster, and starts degrading in performance. The difference in sum rate balloons even more after the second clustering interval. Our proposed technique, alternatively, trains for clusters of different sizes. Therefore, if a single user diverges its path too far from the rest of the users, one UAV can be re-tasked to serve it individually while the remaining two UAVs serve the other remaining 5 users. Overall, our model yields a 9\% increased throughput for the service period as compared to the mutual learning algorithm. This figure demonstrates that our scheme fits more for scenarios with increased service periods and when the users diverge. The benchmark, however, does perform better during short intervals and when the users are in close vicinity to one another.

Fig. \ref{Throughput_vs_Maximum_User_Speed} exhibits the variation in throughput of test episode against maximum user speed, $V_{max}$, for different values for transmitted power, $P_{max}$. Three and four UAVs are employed to serve 6 and 10 users, respectively. The minimum data rate threshold for QoS is selected as per $P_{max}$. The results show that $P_{max}$ has little bearing on data rates for low vehicular speeds. This may be due to the fact that vehicles remain relatively close together for most periods and hence suffer from high inter-cluster interference at higher $P_{max}$. Throughput then increases to a max before starting to decline at higher user speeds, in which the users outrun the UAVs. This decline of throughput at higher user speeds is more prominent at lower $P_{max}$ since UAVs with higher $P_{max}$ have an extended range of communication and are able to communicate with some degree of efficiency even when the users are relatively distant. The figure also shows that the solution scales well to increasing numbers of UAVs and users.

Fig. \ref{Throughput_vs_trajectory_comparison} compares the proposed technique with various other trajectories, including the mutual learning scheme and circular trajectory. Without training, the circular trajectory performs better than other techniques as it covers a pre-defined radius and can serve users as time progresses. DQN trains agents independently and converges slowly. SDQN-2D converges faster than the proposed technique but loses the control over height and displays lower final throughput. Mutual Learning scheme \cite{9520121} clusters users equally amongst UAVs and hence has a smaller action space and converging faster than the proposed scheme. This setting, however, forces sub-optimal clustering and therefore results in lower final throughput. SDQN-3D without action masking allows the UAVs to take irrelevant actions, for which the system penalizes the achieved reward. The SDQN-3D yields a 9\% and 20\% improvement as compared to Mutual learning and separate DQN schemes, respectively, but converges 26\% slower as compared to Mutual Learning scheme.

\section{Conclusion}
This research is motivated by seeking to overcome existing limitations when employing multi-agent RL for multiple UAVs in NOMA-enabled communications networks - especially to address the question of shared learning for agents with different action spaces. We aim to maximize system sum rate by jointly optimising the 3D trajectory of a UAV fleet and power allocation for mobile ground users using a novel SDQN algorithm. A multi-faceted evaluation of our results was undertaken by comparing it with contemporary techniques in terms of convergence time, trajectories, multiple access scheme, ground user speed, learning rates and impact of dynamic re-clustering. The results show that proposed solution outperforms the relevant benchmark implementations in terms of throughput (9\% increase over \cite{9520121}) and convergence time (10\% decrease over seperately trained DQNs). The proposed solution is suitable for settings that require multiple agents to be trained with partially or non-overlapping action spaces. A key limitation of our solution is that it does not take UAV flight energy into account in the reward function and, therefore, exhibits a jittery trajectory. This presents an opportunity for future work in terms of energy efficiency trajectory design. 
\balance

\end{document}